\documentclass[preprint2]{aastex}
\usepackage{color}
\usepackage{epsfig}
\usepackage[fleqn]{amsmath}
\usepackage{multirow}
\usepackage{mathptmx}
\usepackage{graphicx}
\usepackage{bm}

\usepackage{lineno}
\shorttitle{Explicit symplectic methods} \shortauthors{Wu et al.}

\begin{document}


\title{Explicit symplectic methods in black hole spacetimes}


\author{Xin Wu$^{1-3,\dag}$, Ying Wang$^{1,2}$, Wei Sun$^{1,2}$,
Fu-Yao Liu$^{1}$, Wen-Biao Han$^{4-7}$} \affil{1. School of
Mathematics, Physics and Statistics, Shanghai University of
Engineering Science, Shanghai 201620, China
\\ 2. Center of Application and Research of Computational Physics,
Shanghai University of Engineering Science, Shanghai 201620, China
\\  3. Guangxi Key Laboratory for Relativistic Astrophysics, Guangxi
University, Nanning 530004, China  \\  4. Shanghai Astronomical
Observatory, Chinese Academy of Sciences, Shanghai 200030, China
\\ 5. Hangzhou Institute for Advanced Study, University of Chinese Academy of Sciences, Hangzhou 310124, China
\\ 6. School of Astronomy and Space Science, University of Chinese Academy of Sciences, Beijing 100049, China \\
7. Shanghai Frontiers Science Center for Gravitational Wave
Detection, 800 Dongchuan Road, Shanghai 200240, China}
\email{Emails:  $\dag$ Corresponding Author:
wuxin$\_$1134@sina.com (X. W.); wangying424524@163.com (Y. W.),
sunweiay@163.com (W. S.), liufuyao2017@163.com (F. L.),
wbhan@shao.ac.cn (W. H.)}
%\email{wangying424524@163.com} sunweiay@163.com, liufuyao2017@163.com


\begin{abstract}

Many Hamiltonian problems in the Solar System are separable or
separate into two analytically solvable parts, and thus give a
great chance to the development and application of explicit
symplectic integrators based on operator splitting and composing.
However, such constructions cannot in general be available for
curved spacetimes in general relativity and modified theories of
gravity, because these curved spacetimes correspond to
nonseparable Hamiltonians without the two part splits. Recently,
several black hole spacetimes such as the Schwarzschild black hole
were found to allow the construction of explicit symplectic
integrators, since their corresponding Hamiltonians are separable
into more than two explicitly integrable pieces. Although some
other curved spacetimes including the Kerr black hole do not exist
such multi part splits, their corresponding appropriate time
transformation Hamiltonians do. In fact, the key problem for the
obtainment of symplectic analytically integrable decomposition
algorithms is how to split these Hamiltonians or time
transformation Hamiltonians. Considering this idea, we develop
explicit sympelcetic schemes in curved spacetimes. We introduce a
class of spacetimes whose Hamiltonians are directly split into
several explicitly integrable terms. For example, the Hamiltonian
of rotating black ring has a 13 part split. We also present two
sets of spacetimes whose appropriate time transformation
Hamiltonians have the desirable splits. For instance, an 8 part
split exists in a time-transformed Hamiltonian of Kerr-Newman
solution with disformal parameter. In this way, the proposed
symplectic splitting methods will be used widely for long-term
integrations of orbits in most curved spacetimes we have known.

\end{abstract}


\emph{Unified Astronomy Thesaurus concepts}: Black hole physics
(159); Computational methods (1965); Computational astronomy
(293); Celestial mechanics (211)

%\keywords{Computational methods --- Computational astronomy}



\section{Introduction}
\label{sec:intro}

Symplectic integration methods (Hairer et al. 1999; Feng $\&$ Qin
2009) preserve the phase space structure of Hamiltonian dynamics
and do not cause certain constants of motion (e.g., energy) to
have unphysical drifts over large time spans. They yield numerical
solutions, which inherit the qualitative properties of the exact
solutions. Because of these good properties, symplectic
integrators are widely used for long-term numerical integrations
of various dynamical evolution problems in molecular dynamics,
quantum and celestial mechanics.

Near integrable Hamiltonian systems involving planetary N-body
problems in the Solar System are  separable to the variables or
can be split into two analytically solvable parts in general.
Hence, symplectic analytically integrable decomposition
algorithms, as explicit symplectic integration  algorithms based
on splits and compositions, are easily available. The second-order
leapfrog splitting method of Wisdom $\&$ Holman (1991) is one of
the most efficient symplectic methods for the long term numerical
simulation of planetary dynamics in Jacobi coordinates. High order
methods (Yoshida 1990; Chambers $\&$  Murison 2000; Blanes $\&$
Moan 2002) can be developed by composing the exact flows of the
two parts. Besides the two part split, multi part splits have been
applied to the construction of symplectic analytically integrable
decomposition algorithms for Hamiltonian systems that split in
more than two parts (Malhotra 1991; Duncan et al. 1998; Levision
$\&$ Duncan 2000; Wu  et al. 2003). High order multi part split
symplectic integrators by composing many different operators have
also appeared in the literature (Blanes et al. 2008, 2010; Skokos
et al. 2014).

Although geodesic orbits in several standard general relativity
curved spacetime backgrounds such as a Schwarzschild metric and  a
Kerr metric are integrable, the known integrability only shows the
solutions in terms of quadratures rather than those in terms of
elementary functions. Numerical integration schemes are essential
to study these geodesics. If magnetic fields as extra sources are
included in the curved spacetimes, the  orbits  become more
complicated and are even nonintegrable and chaotic in many
circumstances. Numerical integrations are more important to solve
such nonintegrable orbits. In general, the motions of photons or
test particles in general relativity or modified theories of
gravity can be described in terms of Hamiltonian systems,
therefore, symplectic methods are suitable for use without doubt.

The Hamiltonians obtained from curved spacetimes are nonseparable,
or cannot be decomposed onto two explicitly integrable pieces.
This had arisen an obstacle to the implementation of symplectic
analytically integrable decomposition algorithms for a long time.
Instead, implicit symplectic methods were occasionally met in some
literatures on relativistic astrophysics. The implicit midpoint
rule was regarded as a variational-symplectic integrator for
application to general relativity and other constrained
Hamiltonian systems (Brown 2006). This variational integrator was
developed to solve general nonconservative systems (Tsang et al.
2015).  The implicit Gauss-Legendre Runge-Kutta symplectic method
was employed to detect a transition from regular to chaotic
circulation in magnetized coronae near rotating black holes
(Kop\'{a}\v{c}ek et al. 2010). Symmetric, symplectic
Gauss-Runge-Kutta collocation methods with step size controllers
were used to  integrate geodesic orbits in spacetime backgrounds
corresponding to nonintegrable Hamiltonian systems (Seyrich $\&$
Lukes-Gerakopoulos 2012). These integrators preserve the
symplectic form, conserve Noether charges, and exhibit excellent
long-term energy behavior. They are implicit and then are
numerically more expensive to solve than explicit integration
schemes. They are directly applied to Hamiltonian systems that do
not need any splits, and all phase space variables are completely,
implicitly solved. In this sense, they belong to completely
implicit algorithms. Nevertheless, splitting and composition
methods are used in some implicit integration schemes.  In the
context of a splitting of a Hamiltonian into two or more parts,
some individual parts have explicit solutions obtained from
analytical methods or an explicit leapfrog integrator, while the
others have implicit solutions given by the implicit midpoint
rule. By the composition of the explicit solutions of the
subsystems and the implicit solutions of the other subsystems,
explicit and implicit mixed symplectic splitting integrators are
obtained. An explicit and implicit mixed symplectic integrator
with adaptive time steps was used to calculate post-Newtonian
effects of the Kerr metric in the Galactic center region (Preto
$\&$ Saha 2009). Lubich et al. (2010) composed a noncanonically,
explicit and implicit mixed symplectic integration scheme for a
post-Newtonian Hamiltonian of a spinning black-hole binary. The
method is based on a splitting of the Hamiltonian into an orbital
contribution with numerical solutions, and two spin (spin-orbit,
and spin-spin) contributions with analytical solutions. Such an
integrator can become canonical when the conjugate spin
coordinates of Wu $\&$ Xie (2010) are adopted. More intensive
studies on this topic were given by Zhong et al. (2010), Mei et
al. (2013a), and Mei et al. (2013b). These explicit and implicit
mixed symplectic splitting methods should be numerically less
expensive to solve than the completely implicit symplectic
nonsplitting algorithms.

Considering the superiority of explicit integrators in
computational efficiency, several authors have attempted to
develop explicit methods for nonseparable Hamiltonian systems like
those in curved spacetimes. Chin (2009) designed explicit
symplectic integrators for a \emph{selected} class of nonseparable
Hamiltonians, which are product forms of functions with respect to
momenta and functions vs position coordinates. Although these
explicit integrators do not need any splits of the Hamiltonians,
their applications are limited to only the selected Hamiltonians.
To present explicit leapfrog splitting methods for an inseparable
Hamiltonian system, Pihajoki (2015) obtained an extended phase
space new Hamiltonian, which is the sum of the original
Hamiltonian depending on the original  momenta and new coordinates
and another identical copy depending on the original coordinates
and new momenta. Clearly, the newly extended Hamiltonian has a two
part split although the original Hamiltonian is inseparable. The
two part split leapfrog method shows good long term stability and
error behaviour. However, it is not symplectic in the original
phase space and the extended phase space  because it is combined
with coordinate mixing transformations. The phase space mixing
maps were improved by Liu et al. (2016) and Luo et al. (2017). In
particular, the midpoint permutations between the coordinates and
those between the momenta were regarded as the best choice of the
phase space mixing maps (Luo et al. 2017; Li $\&$ Wu 2017; Liu
$\&$ Wu 2017; Wu $\&$ Wu 2018). Tao (2016) did not use any mixing
maps and established three part split explicit  methods for
nonseparable Hamiltonians in extended phase spaces. These
algorithms are symplectic in the extended phase space but are not
in the original phase space. Jayawardana $\&$ Ohsawa (2022) and
Ohsawa (2022) proposed semiexplicit symplectic integrators for
nonseparable Hamiltonian systems. This method, as a combination of
explicit methods and implicit ones, is symplectic in the original
phase space and the extended phase space.

Recently, multi part split methods were applied to several
\emph{individual} curved spacetimes so as to successfully
construct explicit symplectic analytically integrable
decomposition algorithms based on splitting and composition.
Splitting the Hamiltonian for the description of charged particles
moving near a Schwarzschild black hole with an external magnetic
field into four terms, Wang et al. (2021a) designed four part
split explicit symplectic integrators. A Reissner-Nordstr\"{o}m
black hole corresponds to a Hamiltonian separable into 5 terms and
allows for the use of explicit symplectic methods (Wang et al.
2021b). The Hamiltonian describing charged particles moving near a
magnetized Reissner-Nordstr\"{o}m anti-de Sitter black hole
separable into 6 terms is required (Wang et al. 2021c). McLachlan
(2022) showed that such multi part split methods in these curved
spacetimes are well appropriate for application to high order
symplectic partitioned Runge-Kutta and Runge-Kutta-Nystr\"{o}m
optimized methods of Blanes and Moan (2002). Zhou et al. (2022)
claimed that the splitting methods of the Hamiltonians associated
to curved spacetimes  are not unique but have various options. In
addition, the number of splitting pieces should be as small as
possible so that roundoff errors are reduced. However, such multi
part splits are not applicable to the Hamiltonian of the Kerr
metric. Wu et al. (2021) found that an appropriate
time-transformed Hamiltonian has five splitting parts and allows
the construction of explicit symplectic integrators. The five part
split is also suited for the Hamiltonian for the description of
charged particles moving near the Kerr black hole (Sun et al.
2021a).

Are there any other curved spacetimes allowing for the application
of explicit symplectic integrators besides the above-mentioned
individual curved spacetimes?  Which Hamiltonians of curved
spacetimes have multi part splits? Which Hamiltonians of curved
spacetimes do not have but appropriate time transformation
Hamiltonians of curved spacetimes have? To solve these problems,
we shall introduce a class of  curved spacetimes which correspond
to Hamiltonians with multi part splits and two sets of curved
spacetimes which correspond to time transformation Hamiltonians
with multi part splits. Such a great extension to the application
of explicit symplectic integrators in curved spacetimes is the
main aim of this paper.

The remainder of this paper is organized as follows. In Section 2,
we briefly introduce  symplectic splitting and composition methods
for a Hamiltonian with multi split parts in the literature. In
Section 3, we demonstrate how to directly split Hamiltonians in a
class of curved spacetimes. In Section 4, we provide two sets of
curved spacetimes whose corresponding Hamiltonians are not
directly split in several explicitly integrable pieces but
time-transformed Hamiltonians are. Finally, the main results are
concluded in Section 5.

\section{Symplectic splitting and composition methods}

Splitting and composition methods are a main path for the
obtainment of explicit symplectic integrators. Suppose a
$2n$-dimensional Hamiltonian system with $n$-dimensional momentum
$\textbf{p}$ and $n$-dimensional coordinate $\textbf{q}$ is
decomposed into many pieces:
\begin{equation}
H(\textbf{p},\textbf{q})=\sum^{k}_{i=1}H_i(\textbf{p},\textbf{q}),
\end{equation}
where all sub-Hamiltonians $H_i$ can be integrated exactly and
have analytical solutions as explicit functions of time. A series
of operators $\varphi_i$ are analytical solvers of the
sub-Hamiltonians $H_i$. $\varphi_i$ are symplectic operators, and
Equation (1) is a symplectic splitting method of the total
Hamiltonian $H$. The exact solution of each of the
sub-Hamiltonians from the starting solution
$\textbf{z}_0=(\textbf{p}(0),\textbf{q}(0))$ through a time step
$h$ is expressed as
$\textbf{z}=(\textbf{p}(h),\textbf{q}(h))=\varphi^h_i
(\textbf{z}_0)$.

Combining these solutions produces a first-order approximation to
the exact solution of the Hamiltonian system $H$:
\begin{equation}
\chi_h=\varphi^h_k\times\cdots\times\varphi^h_1.
\end{equation}
Its adjoint reads
\begin{equation}
\chi^{*}_{h}=\varphi^h_1\times\cdots\times\varphi^h_k.
\end{equation}
The ordering of terms in the two flow operators may affect the
accuracy of the two flows (McLachlan 2022). The two operators can
symmetrically compose a second-order explicit symplectic scheme
for $H$:
\begin{equation}
S_2(h)=\chi_{h/2}\times\chi^*_{h/2}.
\end{equation}
That is, the Hamiltonian system $H$ has a second-order
approximation solution $\textbf{z}=S_2(h,\textbf{z}_0)$.
Increasing the order of such an integrator by composition yields a
fourth-order explicit symplectic method of Yoshida (1990) as
follows:
\begin{eqnarray}
    S_4=S_2(h\gamma_1)\times S_2(h\gamma_2)\times S_2(h\gamma_1),
\end{eqnarray}
where $\gamma_1 =1/(1-\sqrt[3]{2})$ and $\gamma_2 =1-2\gamma_1$.
An optimal fourth-order  explicit symplectic
Runge-Kutta-Nystr\"{o}m (RKN) method can also be obtained by a
symmetric composition of the two operators $\chi$ and $\chi^*$. It
is expressed as
\begin{eqnarray}
  RKN_{6}4 &=& \chi_{h\alpha_{12}}\times \chi^*_{h\alpha_{11}} \times \chi_{h\alpha_{10}}\times
  \chi^*_{h\alpha_{9}} \nonumber \\
  &&  \times \chi_{h\alpha_{8}}\times
  \chi^*_{h\alpha_{7}} \times \chi_{h\alpha_{6}} \times
  \chi^*_{h\alpha_{5}}  \\
  && \times \chi_{h\alpha_{4}} \times
  \chi^*_{h\alpha_{3}} \times \chi_{h\alpha_{2}} \times
  \chi^*_{h\alpha_{1}}. \nonumber
\end{eqnarray}
The related time coefficients for $k=2$ in the Hamiltonian (1)
were given by Blanes $\&$ Moan (2002). For  $k>2$ in the
Hamiltonian (1), Zhou et al. (2022) gave the time coefficients of
Equation (6):
\begin{eqnarray}
&& \alpha_{1}=\alpha_{12}=0.082984402775764, \nonumber \\
&& \alpha_{2}=\alpha_{11}=0.162314549088478, \nonumber \\
&& \alpha_{3}=\alpha_{10}=0.233995243906975, \nonumber \\
&&\alpha_{4}=\alpha_{9}=0.370877400040627, \\
&&\alpha_{5}=\alpha_{8}=-0.409933704882860, \nonumber \\
&&\alpha_{6}=\alpha_{7}=0.059762109071016. \nonumber
\end{eqnarray}
Here, the optimization requires that the number of the time
coefficients should be more than that of the order conditions and
the coefficients should be determined by minimizing the sum of the
square of coefficients of the fifth-order truncation error terms.
In addition, several high order three part split symplectic
integrators were presented by Skokos et al. (2014).

Clearly, such splitting and composition methods for the
construction of explicit symplectic integrators acting on a
Hamiltonian consist of three steps. They need splitting the
Hamiltonian into two or more pieces in an  appropriate way,
solving an exactly analytical solution of each piece\footnote{The
splitting Hamiltonian method is said to be appropriate only when
the analytical solution is an explicit function of time.}, and
combining these solutions to construct various approximations for
the Hamiltonian. These splitting methods belong to an important
class of geometric numerical integrators, which preserve
structural properties\footnote{These structural features involve
symplecticity, volume, time-symmetry and first integrals.} of the
exact solution. In what follows, we consider the application of
splitting and composition methods to curved spacetimes.

\section{Direct splitting methods in a set of curved spacetimes}

At first we provide a family of curved spacetimes, whose
Hamiltonians can be directly split in the form (1) so as to allow
for the application of explicit symplectic integrators like
Equations (4)-(6). Then, we list several examples on this kind of
spacetime metrics.

\subsection{A class of curved spacetimes}

The number $k=2$ corresponding to the splitting Hamiltonian pieces
in Equation (1) is suitable for numerous Newtonian gravitational
problems in the solar system. However, it is not in general for
relativistic gravitational problems in curved spacetimes.
Recently, our group found in a series of works that the splitting
forms with $k>2$ are probably admissible in curved spacetimes. The
Hamiltonian associated to the Schwarzschild spacetime can be
decomposed into four (i.e., $k=4$) integrable parts having
analytical solutions as explicit functions of proper time (Wang et
al. 2021a). It also accepts the  number of splitting terms $k=3$
(Zhou et al. 2022). The splitting number is $k=5$ for the
Reissner-Nordstr\"{o}m black hole (Wang et al. 2021b) and a
magnetized modified gravity Schwarzschild spacetime (Yang et al.
2022). The splitting number is $k=6$ for the
Reissner-Nordstr\"{o}m-(anti)-de Sitter black hole (Wang et al.
2021c).  These splitting methods in the three examples are
dependent on concrete black hole metrics. Which black hole metrics
have the direct splitting forms of Equation (1)? Let us seek a set
of universal spacetime metrics meeting this requirement.

Setting $x^{\mu}=(t,u,v,w)$ as spacetime coordinates, we consider
a generic spacetime metric
\begin{eqnarray}
ds^2 &=& -f_0(u,v)dt^2 +2f_{03}(u,v)dtdw + f_3(u,v)dw^2 \nonumber
\\ && +\frac{f_{11}(v)}{f_{12}(u)}du^2+\frac{f_{21}(u)}{f_{22}(v)}dv^2.
\end{eqnarray}
Here, $f_0$, $f_{03}$ and $f_3$ are functions of $u$ and $v$;
$f_{11}$ is a function of $v$, and $f_{21}$ is a function of $u$.
Functions $f_{12}$ and $f_{22}$ are supposed to have the
expressions
\begin{eqnarray}
f_{12} &=& \sum^{j_1}_{i=0} b_iu^{a_i}+ \sum^{l_1}_{i=0}
d_i(u+\kappa_1)^{c_i}, \\
f_{22} &=& \sum^{j_2}_{i=0} \bar{b}_iv^{\bar{a}_i}+
\sum^{l_2}_{i=0} \bar{d}_i(v+\kappa_2)^{\bar{c}_i},
\end{eqnarray}
where $a_i$, $b_i$, $c_i$, $d_i$, $\bar{a}_i$, $\bar{b}_i$,
$\bar{c}_i$, $\bar{d}_i$, $\kappa_1$ and $\kappa_2$ are constant
parameters expressed in terms of real numbers. This metric
corresponds to the Lagrangian formulism
\begin{eqnarray}
\mathcal{L} &=& \frac{1}{2}\frac{ds^2}{d\tau^2} \nonumber
\\
&=& -\frac{1}{2}f_0(u,v)\dot{t}^2 +f_{03}(u,v)\dot{t}\dot{w} +
\frac{1}{2}f_3(u,v)\dot{w}^2 \nonumber
\\ && +\frac{1}{2}
\frac{f_{11}(v)}{f_{12}(u)}\dot{u}^2
+\frac{1}{2}\frac{f_{21}(u)}{f_{22}(v)}\dot{v}^2,
\end{eqnarray}
where 4-velocities $(\dot{t},\dot{u},\dot{v},\dot{w})$ are
derivatives of spacetime coordinates $x^{\mu}=(t,u,v,w)$ with
respect to proper time $\tau$. Based on the Lagrangian
$\mathcal{L}$, generalized momenta are defined as
$p_{x^{\mu}}=\partial\mathcal{L}/\partial \dot{x}^{\mu}$, that is,
\begin{eqnarray}
p_t &=& -f_0(u,v)\dot{t}+f_{03}(u,v)\dot{w}=-E, \\
 p_u &=&
\frac{f_{11}(v)}{f_{12}(u)}\dot{u}, \\
 p_v &=& \frac{f_{21}(u)}{f_{22}(v)}\dot{v}, \\
 p_w &=& f_3(u,v)\dot{w}+f_{03}(u,v)\dot{t}=L.
\end{eqnarray}
$E$ is a  conserved energy of a test particle moving the
gravitational field, and $L$ is also a constant of motion of a
test particle. This Lagrangian is exactly equivalent to the
Hamiltonian formulism
\begin{eqnarray}
H &=&
\frac{1}{2}(\frac{L^2}{f_3}-\frac{E^2}{f_0})+f_{03}(Ef_3+Lf_{03})\frac{(Lf_0-Ef_{03})}{(f_0f_3+f^2_{03})^2}
\nonumber
\\ && +\frac{1}{2} \frac{f_{12}(u)}{f_{11}(v)}p^2_u
+\frac{1}{2}\frac{f_{22}(v)}{f_{21}(u)}p^2_v.
\end{eqnarray}
The Hamiltonian has two degrees of freedom and a four-dimensional
phase space. If the particle is time-like, the Hamiltonian is
always identical to a given constant
\begin{eqnarray}
H =-\frac{1}{2},
\end{eqnarray}
because the 4-velocities satisfy the relation
$\dot{x}^{\mu}\dot{x}_{\mu}=-1$. Here, the speed of light is taken
as one geometric unit, $c=1$. The constant of gravity  also uses
one geometric unit, $G=1$. Now, the Hamiltonian (16) can be
directly separated in the form
\begin{eqnarray}
H &=& H_1+ \sum^{j_1}_{i=0} H_{b_i}+ \sum^{l_1}_{i=0}
H_{d_i} \nonumber \\
&& +\sum^{j_2}_{i=0}H_{\bar{b}_i}+ \sum^{l_2}_{i=0} H_{\bar{d}_i};
\\
H_1 &=& f_{03}
(Ef_3+Lf_{03})\frac{(Lf_0-Ef_{03})}{(f_0f_3+f^2_{03})^2} \nonumber \\
&& +\frac{1}{2}(\frac{L^2}{f_3}-\frac{E^2}{f_0}), \\
H_{b_i} &=& \frac{1}{2} \frac{b_iu^{a_i}}{f_{11}(v)}p^2_u , \\
H_{d_i} &=& \frac{1}{2} \frac{d_i(u+\kappa_1)^{c_i}}{f_{11}(v)}p^2_u , \\
H_{\bar{b}_i} &=& \frac{1}{2}\frac{\bar{b}_iv^{\bar{a}_i}}{f_{21}(u)}p^2_v , \\
H_{\bar{d}_i} &=&
\frac{1}{2}\frac{\bar{d}_i(v+\kappa_2)^{\bar{c}_i}}
{f_{21}(u)}p^2_v.
\end{eqnarray}
Obviously, each of the sub-Hamiltonians (19)-(23) is analytically
solvable and its solutions are  explicit functions of proper time
$\tau$. In other words, the Hamiltonian (18) resembles Equation
(1), where $\textbf{q}=(u,v)$ and $\textbf{p}=(p_u,p_v)$. Thus,
the explicit symplectic integrators such as Equations (4)-(6) are
applicable to the spacetime metric (8).

Consider that an asymptotically uniform electromagnetic field
exists in the vicinity of the central body. This electromagnetic
field is assumed to have a four-vector potential with two nonzero
components $A_t$ and $A_w$ as functions of $u$ and $v$. The motion
of a test particle with charge $e$ around the central body is
represented by the Hamiltonian
\begin{eqnarray}
H_e &=&
\frac{(L-eA_w)^2}{2f_3}-\frac{(E+eA_t)^2}{2f_0}+f_{03}[(L-eA_w)f_{03}
\nonumber
\\ && -(E+eA_t)f_3] [(L-eA_w)f_0+(E+eA_t)f_{03}] \nonumber
\\ && \times(f_0f_3+f^2_{03})^{-2} +\frac{1}{2}
\frac{f_{12}(u)}{f_{11}(v)}p^2_u \nonumber
\\ &&
+\frac{1}{2}\frac{f_{22}(v)}{f_{21}(u)}p^2_v.
\end{eqnarray}
This Hamiltonian still allows for the splitting form (18), where
only minor modifications are given to Equation (19).

Two notable points are given here. Splitting methods in these
curved spacetimes involve four steps: (i) obtaining the
Hamiltonian in terms of the metric; (ii) splitting the
Hamiltonian; (iii) exactly solving each of the splitting pieces;
and (iv) combining these solutions. The black hole spacetimes that
allow such splitting methods in the previous studies (Wang et al.
2021a, 2021b, 2021c) are several examples of the metric family
(8).  A modified gravity Schwarzschild black hole solution based
on  the scalar-tensor-vector modified gravitational theory (Yang
et al. 2022) resembles one of the metric family (8). Brane-world
black holes (Deng 2020a; Hu $\&$ Huang 2022) are also an example
of the metric family (8). It was shown in the previous works that
the explicit symplectic integrators have an advantage over the
implicit symplectic methods, and the implicit and explicit mixed
symplectic methods at the same order in computational efficiency.
The explicit integrators have such good computational efficiency
regardless of the type of Hamiltonian systems. Besides of these
mentioned spacetimes, other black hole metrics belonging to the
metric family (8) are present. Two of them are listed in what
follows.

\subsection{Reissner-Nordstr\"{o}m spacetime with extra sources}

Boyer-Lindquist coordinates $(t,r,\theta,\phi)$ corresponding to
the spacetime coordinates $(t,u,v,w)$ in Eq. (8) are chosen. In
this coordinate system, a spherically-symmetric static
Reissner-Nordstr\"{o}m-(de Sitter)-Anti-de Sitter black hole
surrounded by extra sources such as quintessence and a cloud of
strings has a covariant metric (Kiselev 2003)
\begin{equation}
ds^{2}
=g_{tt}dt^{2}+g_{rr}dr^{2}+g_{\theta\theta}d\theta^{2}+g_{\phi\phi}d\phi^{2},
\end{equation}
where four nonzero metric components are
\begin{eqnarray}
g_{tt} &=& -f_0(r), \\
g_{rr} &=& \frac{1}{f_0(r)}, \\
g_{\theta\theta} &=& r^2,   \\
g_{\phi\phi} &=& r^{2}\textrm{sin}^{2}\theta.
\end{eqnarray}
Function $f_0(r)$ is expressed as
\begin{eqnarray}
f_0(r) = (1-b_c-\frac{2M}{r}+\frac{Q^2}{r^2} -\frac{\Lambda}{3}r^2
-\frac{\alpha_q}{r^{3\omega_q+1}}).
\end{eqnarray}
The related notations in Equation (30) are given below.

$M$ and $Q$ are the mass and charge of the black hole. $\Lambda$
denotes a cosmological constant. The cosmological constant
associated with the vacuum energy can provide a negative pressure
responsible for the accelerating expansion of the Universe
(Perlmutter et al. 1999). $\omega_{q}$ represents a quintessential
state parameter, and $\alpha_{q}$ stands for a quintessence
parameter. The quintessence parameter equating to the ratio of the
pressure and density is the so-called quintessential state
equation. The quintessential state parameter can characterize a
dark energy and therefore is termed the quintessential dark
energy. The quintessence is regarded as another of the origins of
the negative pressure causing the accelerating expansion of the
Universe. The quintessence field demands $\alpha_{q}>0$ and
$\omega_{q}<0$. $\omega_{q}=-1$ plays a role of cosmological
constant. The ranges of quintessential state parameter are
$\omega_{q}<-1$ for the phantom energy and $-1<\omega_{q}<-1/3$
for the quintessence. Obviously, the existence of the vacuum
energy or the quintessence changes the asymptotic structure of
black hole, but still allows for the presence of cosmological
horizons. In fact, the quintessence field is obtained via the
Einstein gravity coupled to a scalar field, and is an alternative
or extension of the standard Einstein gravity. See the paper of
Toledo $\&$ Bezerra (2020) for more information on the
quintessence matter surrounding a black hole. In addition, $b_c$
is a parameter for measuring the intensity of a cloud of strings
around the black hole. The cloud formed by strings can be viewed
as a source of the gravitational field, where the Universe is
described by a collection of extended objects corresponding to
one-dimensional strings, but is not represented by a collection of
point particles (Letelier 1979). In short, Equation (30) includes
all these different gravitational sources, like the cosmological
constant, the quintessence matter and the cloud of strings. The
obtainment of Equation (30) is based on the assumption that the
energy-momentum tensor is a linear superposition of the
energy-momentum tensors associated with each one of the sources.

The black hole metric (25) as one of the metric family (8)
corresponds to the Hamiltonian
\begin{equation}
H=\sum^{7}_{i=1}H_i,
\end{equation}
where these sub-Hamiltonian parts are
\begin{eqnarray}
H_1 &=&-\frac{E^2}{2f_0(r)}+\frac{L^2}{2r^2\sin^2\theta}, \\
H_2 &=& \frac{1-b_c}{2}p^2_r, \\
H_3 &=& -\frac{M}{r}p^2_r, \\
H_4 &=& \frac{Q^2}{2r^2}p^2_r, \\
H_5 &=&  -\frac{\Lambda}{6}r^2p^2_r, \\
H_6 &=&  -\frac{\alpha_q p^2_r}{2r^{3\omega_q+1}}, \\
H_7 &=&  \frac{p^2_\theta}{2r^2}.
\end{eqnarray}
$L$ in Equation (32) stands for the particle's angular momentum.
Equations (33)-(37) stem from the splittings of the term
$f_0(r)p^2_r/2$. The seven pieces exist their analytical solutions
as explicit functions of proper time $\tau$. According to the
result of Cao et al. (2022), the sum of the seven splitting
pieces, i.e. the Hamiltonian $H$ of Equation (31), is integrable.
Notice that each of the seven splitting pieces is integrable or
not, irrespective of whether the sum of the seven splitting pieces
is integrable. Each piece is integrable, but the sum may be
nonintegrable. Now, explicit symplectic algorithms like Equations
(4)-(6) can work in the metric (25).

The splitting method with several explicitly integrable pieces is
not unique, as was claimed in the work of Zhou et al. (2022). For
instance, the number of splitting pieces in Equation (31) is six
when the sum of $H_2$ and $H_7$ is considered.

\subsection{Rotating black ring}

Emparan $\&$ Reall (2002) gave a solution of the vacuum Einstein
equations in five dimensions to a rotating black ring. In ring
coordinates $(t,x,y,\phi,\psi)$ with $|x|\leq 1$ and $y\leq -1$,
the solution was written in the paper of Igata et al. (2011) as
\begin{eqnarray}
ds^2 &=&
-\frac{F(y)}{F(x)}\left(dt-CR\frac{1+y}{F(y)}d\psi\right)^2
 \nonumber \\
&& +\frac{R^2F(x)}{(x-y)^2}
(-\frac{G(y)}{F(y)}d\psi^2-\frac{dy^2}{G(y)}
\nonumber \\
&& +\frac{dx^2}{G(x)}+\frac{G(x)}{F(x)}d\phi^2).
\end{eqnarray}
$R>0$ is a parameter representing the black ring's radius. $C$ is
also a parameter characterizing the rotation velocity $\lambda$
and the thickness $\nu$ of the ring, and is expressed as
$C=[\lambda(\lambda-\nu)(1+\lambda)/(1-\lambda)]^{1/2}$ with
$0<\nu\leq\lambda<1$. Two functions are
\begin{eqnarray}
F(z)=1+\lambda z, ~~~~ G(z)=(1-z^2)(1+\nu z).
\end{eqnarray}
The black ring  metric is stationary asymptotically flat and has
an event horizon of non-spherical topology. Although $y=-1/\nu$ is
the position of the event horizon, it is not when the polar
coordinates $(y, \psi)$ are transformed into Cartesian
coordinates. Regularity of the full metric at the ring axis and
the equatorial plane exists for the condition
$\lambda=2\nu/(1+\nu^2)$. This means that the spacetime can be
completely regular on and outside the event horizon of
non-spherical topology in this case.

The black ring  metric (39) exactly corresponds to the Hamiltonian
\begin{eqnarray}
H &=& H_1+\frac{1}{2}g^{xx}p^2_x+\frac{1}{2}g^{yy}p^2_y,
\\
H_1 &=& \frac{1}{2}(g^{tt}E^2+g^{\phi\phi}l^2_\phi
+g^{\psi\psi}l^2_\psi  \nonumber \\
&&  -2g^{t\psi}El_\psi),
\end{eqnarray}
where $-E$, $l_\phi$ and $l_\psi$ are constant conjugate momenta,
and these contravariant metric components are
\begin{eqnarray}
g^{tt} &=&
-\frac{F(x)}{F(y)}-\frac{C^2(x-y)^2(y+1)^2}{G(y)F(x)F(y)},
 \\
g^{xx} &=& \frac{(x-y)^2G(x)}{R^2F(x)}, \\
g^{yy} &=& -\frac{(x-y)^2G(y)}{R^2F(x)}, \\
g^{\phi\phi} &=& \frac{(x-y)^2}{R^2G(x)},
\\
g^{\psi\psi} &=& -\frac{F(y)(x-y)^2}{R^2G(y)F(x)},
\\
g^{t\psi} &=& -\frac{C(x-y)^2(y+1)}{R^2G(y)F(x)}.
\end{eqnarray}
The Hamiltonian (41) contains two degrees of freedom and its phase
space has four dimensions. Igata et al. (2011) found that the
black ring geometry does not allow the separation of variables in
the Hamilton-Jacobi equation for Equation (41) but allows the
presence of chaotic bound orbits. This indicates the absence of an
additional constant of motion except the conserved Hamiltonian
(17) and the constants $E$, $l_\phi$ and $l_\psi$ associated with
the Killing vectors. In spit of this, the Hamiltonian (41) exists
a separable form similar to Equation (18).

Splitting  the Hamiltonian (41) requires splitting $g^{xx}$ and
$g^{yy}$ in Equations (44) and (45). Because $g^{yy}$ takes $y$ as
a variable and $x$ as a constant, it is simply split into the form
\begin{eqnarray}
g^{yy} &=& -\frac{1}{R^2F(x)}[x^2+x(\nu x-2)y \nonumber \\
&&  +(1-2\nu x-x^2)y^2+(\nu+2x-\nu x^2)y^3 \nonumber \\
&&  +(2\nu x-1)y^4-\nu y^5].
\end{eqnarray}
The third term of Equation (41) consists of six explicitly
solvable parts
\begin{eqnarray}
H_2 &=& -\frac{x^2 p^2_y }{2R^2F(x)}, \\
H_3 &=& -\frac{xy(\nu x-2)}{2R^2F(x)}p^2_y, \\
H_4 &=& -\frac{y^2 p^2_y}{2R^2F(x)}(1-2\nu x-x^2), \\
H_5 &=& -\frac{y^3 p^2_y}{2R^2 F(x)}(\nu+2x-\nu x^2), \\
H_6 &=& -\frac{y^4 p^2_y}{2R^2F(x)}(2\nu x-1), \\
H_7 &=& \frac{\nu y^5 p^2_y}{2R^2F(x)}.
\end{eqnarray}
As far as $g^{xx}$ is concerned,  $x$ is a variable and $y$ is a
constant. Compared with $g^{yy}$,  $g^{xx}$ has a more complicated
splitting form. Setting $\xi=1+\lambda x$, i.e.,
$x=(\xi-1)/\lambda$, we have
\begin{eqnarray}
g^{xx} = \frac{1}{R^2}\sum^{6}_{i=1} G_i,
\end{eqnarray}
where
\begin{eqnarray}
G_1 &=& -\frac{\nu}{\lambda^5}\xi^4, \\
G_2 &=& (\frac{5\nu}{\lambda^5}+\frac{2\nu y-1}{\lambda^4})\xi^3, \\
G_3 &=& [-\frac{10\nu}{\lambda^5}+\frac{4}{\lambda^4}(1-2\nu y)  \nonumber \\
&&  +\frac{1}{\lambda^3}(\nu+2y-\nu y^2)]\xi^2, \\
G_4 &=& [\frac{10\nu}{\lambda^5}-\frac{6}{\lambda^4}(1-2\nu
y)-\frac{3}{\lambda^3}(\nu+2y-\nu y^2) \nonumber \\
&&  +\frac{1}{\lambda^2}(1-2\nu y-y^2)]
\xi, \\
G_5 &=& -\frac{5\nu}{\lambda^5}+\frac{4}{\lambda^4}(1-2\nu
y)+\frac{3}{\lambda^3}(\nu+2y-\nu y^2) \nonumber \\
&&  -\frac{2}{\lambda^2}(1-2\nu y-y^2)+\frac{y}{\lambda}(\nu y-2), \\
G_6 &=& [\frac{\nu}{\lambda^5}-\frac{1}{\lambda^4}(1-2\nu
y)-\frac{1}{\lambda^3}(\nu+2y-\nu y^2) \nonumber \\
&&  +\frac{2}{\lambda^2}(1-2\nu y-y^2) \nonumber \\
&& -\frac{y}{\lambda}(\nu y-2)+y^2]/\xi.
\end{eqnarray}
The second term of Equation (41) has six explicitly solvable parts
\begin{eqnarray}
H_8 &=&    \frac{G_1}{2R^2}p^2_x, \\
H_9 &=&    \frac{G_2}{2R^2}p^2_x, \\
H_{10} &=& \frac{G_3}{2R^2}p^2_x, \\
H_{11} &=& \frac{G_4}{2R^2}p^2_x, \\
H_{12} &=& \frac{G_5}{2R^2}p^2_x, \\
H_{13} &=& \frac{G_6}{2R^2}p^2_x.
\end{eqnarray}
Thus, the Hamiltonian (41) can be split into 13  explicitly
solvable parts
\begin{eqnarray}
H= \sum^{13}_{i=1} H_i.
\end{eqnarray}
Explicit symplectic algorithms like Equations (4)-(6) are
available for the spacetime (39).

In short, the Hamiltonians corresponding to the metric (8) have
the splitting forms  (1). On the other hand, the spacetimes whose
Hamiltonians have such splitting forms are not restricted to the
metric family (8).

\section{Indirect splitting methods in two types of curved spacetimes}

Hamiltonians for some other curved spacetimes like the Kerr metric
are not directly split into Equation (1). However, their
time-transformed Hamiltonians have the splitting form (1), as was
claimed by several authors (Wu et al. 2021; Sun et al. 2021a; Sun
et al. 2021b; Zhang et al. 2021, 2022). In what follows, such two
types of curved spacetimes are given.

\subsection{Type 1: inseparable parts as functions of one variable}

The metric (8) is slightly modified as
\begin{eqnarray}
ds^2 &=& -f_0(u,v)dt^2 +2f_{03}(u,v)dtdw + f_3(u,v)dw^2 \nonumber
\\ && +\frac{f_{11}(v)}{f_{12}(u)e(u)}du^2
+\frac{f_{21}(u)}{f_{22}(v)}dv^2,
\end{eqnarray}
where $f_{12}$ and $f_{22}$ are separable parts given by Equations
(9) and (10), but $e(u)$ is a function of the variable $u$. The
Hamiltonian (16) is also slightly altered as
\begin{eqnarray}
H &=&
\frac{1}{2}(\frac{L^2}{f_3}-\frac{E^2}{f_0})+f_{03}(Ef_3+Lf_{03})\frac{(Lf_0-Ef_{03})}{(f_0f_3+f^2_{03})^2}
\nonumber
\\ && +\frac{e(u)}{2} \frac{f_{12}(u)}{f_{11}(v)}p^2_u
+\frac{1}{2}\frac{f_{22}(v)}{f_{21}(u)}p^2_v.
\end{eqnarray}
Here, the function $e(u)$ is chosen so that the third term of
Equation (71) is inseparable or is not split in the form (1). In
this case, the explicit symplectic integrators (4)-(6) are not
appropriate for the numerical integration of the Hamiltonian (71).
Wu et al. (2021) successfully constructed  the explicit symplectic
methods for the Kerr metric by following the idea of Mikkola
(1997) who introduced time transformation to improve the
efficiency of Wisdom-Holman-like symplectic algorithm for various
hierarchical few-body problems. The time transformed explicit
symplectic algorithms for  the Kerr metric are similarly extended
to the Hamiltonian (71). The implementation of time transformed
explicit symplectic algorithms for the Hamiltonian (71) is briefly
described as follows.

Taking  $\tau=q_0$ as a new coordinate together the corresponding
conjugate momentum $p_0=-H=1/2$, we obtain an extended phase space
$(q_0,u,v,p_0,p_u,p_v)$. A new Hamiltonian in the extended phase
space is
\begin{eqnarray}
\mathcal{H}=g(u)(H+p_0),
\end{eqnarray}
where $g(u)$ is a time transformation function (or a time step
function) from the proper time $\tau$ to a new fictitious time
$\sigma$ in the form
\begin{eqnarray}
d\tau=g(u)d\sigma.
\end{eqnarray}
$\mathcal{H}=0$ for any new time $\sigma$. When the time step
function is chosen as
\begin{eqnarray}
g(u)=\frac{1}{e(u)},
\end{eqnarray}
the Hamiltonian (72) is separable. Its splitting is similar to
Equation (18), but only the differences in Equations (19), (22)
and (23) are as follows: $H_1\rightarrow g(u)(H_1+p_0)$,
$H_{\bar{b}_i}\rightarrow g(u)H_{\bar{b}_i}$ and
$H_{\bar{d}_i}\rightarrow g(u)H_{\bar{d}_i}$. Hence operator
splitting techniques can be used to derive explicit symplectic
integration algorithms like Equations (4)-(6) for the Hamiltonian
(72). These constructions are not directly applicable to the
nonseparable Hamiltonian (71) but act on the separable
time-transformed Hamiltonian (72). They are called indirect
splitting methods. An important role of the time transformation
function $g$ is eliminating the inseparable terms in the
numerators or denominators of the metric functions.

Some notable points are given here. Indirect splitting methods in
these curved spacetimes involve several steps: (i) obtaining the
Hamiltonian in terms of the metric; (ii) extending the phase space
of the Hamiltonian; (iii) finding a time transformation function
that eliminates the inseparable terms in the numerators or
denominators of the metric functions, and writing a
time-transformed Hamiltonian; and (iv) applying the splitting and
composition methods introduced in Section 2 to the
time-transformed Hamiltonian. A constant step-size is used for the
new time $\sigma$ in the proposed algorithms acting on the
time-transformed Hamiltonian (72), but the proper time step will
vary according to Equation (73). The constant step-size can ensure
the good long term behavior of such symplectic methods for the
time-transformed Hamiltonian. The varying time steps are useful to
improve the efficiency of the leap-frog method for various
few-body problems with large eccentricities. If the use of time
transformations is the obtainment of the desirable splitting of
the time-transformed Hamiltonian but is not the consideration of
adaptive time step control to the proposed symplectic integrators,
then specific choices of the time step function are $g(u)\approx
1$. In the next discussions, we list several examples of the
metric family (70).

\subsubsection{Rotating black ring}

We have shown in Section 3.3 that the Hamiltonian (41) without
time transformation is directly split into the 13 explicitly
integrable parts and allow for the construction of the explicit
symplectic integrators. We also use time transformations to
simply establish our algorithms.

One path is the time step function given by
\begin{eqnarray}
g(x)=F(x).
\end{eqnarray}
This leads to  eliminating the function $F(x)$ in Equations (44)
and (45). All the functions $F(x)$ in Equations (50)-(55) are also
eliminated. The second term of Equation (41) is still separated
into 6 explicitly integrable pieces because the numerator of
$g^{xx}$ in Equation (44) is a quintic polynomial of $x$. That is
to say,  the Hamiltonian (41) is still required to have the 13
desirable splitting parts so that it is suitable for the
application of explicit symplectic integrators. A variable proper
time step $\Delta \tau=g(x)\Delta\sigma=hg(x)$ is the range of
$(1-\lambda)h\leq \Delta \tau\leq (1+\lambda)h$.

Another path is the time step function chosen as
\begin{eqnarray}
g(x,y)=-\frac{\nu y^5 F(x)}{G(x)G(y)(x-y)^2}.
\end{eqnarray}
The Hamiltonian (41) corresponds to the time-transformed
Hamiltonian with three analytically solvable parts
\begin{eqnarray}
\mathcal{H} = \mathcal{H}_1+\mathcal{H}_2+\mathcal{H}_3,
\end{eqnarray}
where
\begin{eqnarray}
\mathcal{H}_1 &=& g(x,y)(H_1+p_0), \\
\mathcal{H}_2 &=& -\frac{1}{2}\frac{\nu y^5}{R^2G(y)}p^2_x, \\
\mathcal{H}_3 &=& \frac{1}{2}\frac{\nu y^5}{R^2G(x)}p^2_y.
\end{eqnarray}
This means that $k=3$ in Equations (2) and (3). Thus, the explicit
symplectic methods (4)-(6) are easily available. The choice of the
time step function (76) causes the proper time step $\Delta \tau$
to slightly vary in the vicinity of the fixed new time step $h$.

The above demonstrations show that the two choices of the time
step function yield the explicit symplectic methods. In fact, the
time step function has various choices. A suitable choice of the
time step function can bring a simple construction of  the
algorithms.

\subsubsection{Regular black holes}

Ayon-Beato and Garcia (1998) gave a spherically symmetric black
hole with mass $M$ and charge $Q$ in Schwarzschild coordinates
$(t,r,\theta,\varphi)$:
\begin{eqnarray}
ds^2 = -f(r)dt^2+\frac{dr^2}{f(r)}+r^2(d\theta^2+\sin^2\theta
d\varphi^2),
\end{eqnarray}
where the metric function is
\begin{eqnarray}
f(r) = 1-\frac{2Mr^2}{(r^2+Q^2)^{3/2}}
+\frac{Q^2r^2}{(r^2+Q^2)^{2}}.
\end{eqnarray}
This metric has the event horizon singularity, but lacks curvature
singularities and is regular everywhere. Such a nonsingular metric
solution satisfies the Einstein field equation coupled with
suitable nonlinear electromagnetic fields. That is, it is obtained
from modified or alternative theories of gravity.  It is also
viewed as a Reissner-Nordstr\"{o}m black hole with variable mass
and charge. The metric (81) corresponds to the Hamiltonian
\begin{eqnarray}
H &=& H_1+\frac{f(r)}{2}p^2_r+\frac{1}{2r^2}p^2_\theta, \\
H_1 &=& \frac{1}{2}\left(\frac{L^2}{r^2\sin^2\theta}
-\frac{E^2}{f(r)}\right).
\end{eqnarray}
The second term of Equation (83) has no the desirable splitting
due to the presence of two fractions appearing in $f(r)$. Taking
the time transformation function
\begin{eqnarray}
g(r)=\frac{1}{f(r)},
\end{eqnarray}
we have the time-transformed Hamiltonian
\begin{eqnarray}
\mathcal{H} = \frac{H_1+p_0}{f(r)}+\frac{p^2_r}{2}
+\frac{p^2_\theta}{2r^2f(r)}.
\end{eqnarray}
The three part split is what we want. Zhang et al. (2022) gave a
similar splitting to another regular black hole metric.

There are other regular black holes. Balart $\&$ Vagenas (2014)
found a  regular black hole solution that has the metric (81) with
the metric function (82) being
\begin{eqnarray}
f(r) =1-\frac{2M}{r} \left[\frac{2}{\exp(\frac{Q^2}{4Mr})+1}
\right]^4,
\end{eqnarray}
where $Q=1.153M$. The metric function (87) is an exponential
function and is unlike the metric function (82) being a fractional
function. In spite of this, the time step function obtaining
Equation (85) with Equation (87) still meets the requirement. The
black-bounce-Reissner-Nordstr\"{o}m spacetime (Zhang $\&$ Xie
2022a, 2022b) is globally regular, too.

\subsubsection{Gauss-Bonnet black hole}

The Gauss-Bonnet black hole (Zeng et al. 2020) is a spherically
symmetric black hole whose metric is Equation (81) but metric
function is
\begin{eqnarray}
f(r) =1+\frac{r^2}{2\alpha}\left(1-\sqrt{1+\frac{8\alpha
M}{r^3}}\right),
\end{eqnarray}
where $\alpha$ represents the Gauss-Bonnet coupling constant. Two
horizons exist for $\alpha>0$, while only one horizon exists for
$\alpha<0$. Although the metric function (88) is a radical rather
than a fractional function in equation (82), the same method
induces the time-transformed Hamiltonian resembling Equation (86).

A similar example is Hairy black holes in
Einstein-scalar-Gauss-Bonnet theories (Gao $\&$ Xie 2021). The
Kehagias-Sfetsos asymptotically flat black hole solution of the
modified Ho\v{r}ava-Lifshitz gravity in external magnetic fields
(Abdujabbarov et al. 2011; Stuchl\'{i}k et al. 2014; Toshmatov et
al. 2015) also allows for obtaining the time-transformed
Hamiltonians similar to Equation (86). Some other examples include
4D Einstein-Lovelock black holes (Lin $\&$ Deng 2021),
quantum-corrected Schwarzschild black holes (Deng 2020b; Gao $\&$
Deng 2021; Lu $\&$ Xie 2021), and an Einstein-Lovelock
ultracompact object (Gao $\&$ Xie 2022).

\subsection{Type 2: inseparable parts as functions of two variables}

The metric (8) is slightly modified as
\begin{eqnarray}
ds^2 &=& -f_0(u,v)dt^2 +2f_{03}(u,v)dtdw + f_3(u,v)dw^2 \nonumber
\\ && +j(u,v)\left(\frac{f_{11}(v)}{f_{12}(u)}du^2
+\frac{f_{21}(u)}{f_{22}(v)}dv^2\right),
\end{eqnarray}
where $f_{12}$ and $f_{22}$ are separable parts given by Equations
(9) and (10), but $j(u,v)$ is  a function of the two variables $u$
and $v$ and $1/j(u,v)$ is inseparable.

Taking the time step function
\begin{eqnarray}
g(u,v)=j(u,v)
\end{eqnarray}
derives the time-transformed Hamiltonian (72), which is consistent
with Equation (18) but $H_1$ in Equation (19) should be
$j(u,v)(H_1+p_0)$. Hence, such a time-transformed Hamiltonian
meets the requirement of splitting and composition methods. The
time step function (76) for the rotating black ring is an example
of the time step function (90). Other examples are used to show
the implementation of the algorithms in the following discussions.

\subsubsection{Majumdar-Papapetrou dihole spacetime}

The Majumdar-Papapetrou dihole black holes (Hartle $\&$ Hawking
1972) are two fixed charged black holes in equilibrium under their
gravitational and electrical forces. The Majumdar-Papapetrou
geometry is described in polar coordinates  $(t,\rho,\phi,z)$ by
the metric (Nakashi $\&$ Igata 2019)
\begin{eqnarray}
ds^2 = -\frac{dt^2}{U^2}+U^2(d\rho^2+\rho^2d\phi^2+dz^2),
\end{eqnarray}
where $U$ is a function of $\rho$ and $z$ in the form
\begin{eqnarray}
U = 1+\frac{M_1}{\sqrt{\rho^2+(z-a)^2}}
+\frac{M_2}{\sqrt{\rho^2+(z+a)^2}}.
\end{eqnarray}
$M_1$ and $M_2$ are masses of the two black holes at $z=\pm a$
($a\geq 0$).

Choosing the time step function
\begin{eqnarray}
g(\rho,z)=U^2,
\end{eqnarray}
we obtain the time-transformed Hamiltonian with two splitting
pieces
\begin{eqnarray}
\mathcal{H} &=& \mathcal{H}_1+ \mathcal{H}_2, \\
\mathcal{H}_1 &=& \frac{1}{2}\left(\frac{L^2}{\rho^2}-U^4E^2\right)+p_0U^2, \\
\mathcal{H}_2 &=& \frac{1}{2}\left(p^2_\rho+p^2_z \right).
\end{eqnarray}
Thus, $\chi$ and  $\chi^*$ in Equations (2) and (3) contain two
operators associated with $\mathcal{H}_1$ and $\mathcal{H}_2$ in
Equations (95) and (96). In this way, the explicit symplectic
methods (4)-(6) are applicable to the time-transformed Hamiltonian
$\mathcal{H}$ of Equation (94).

\subsubsection{Reissner-Nordstr\"{o}m-Melvin black holes}

The Reissner-Nordstr\"{o}m-Melvin black holes are a family of
electrovacuum type solutions of the Einstein-Maxwell equations
with scalar field perturbations. They describe the
Reissner-Nordstr\"{o}m black holes permeated by uniform magnetic
fields in the metric (Gibbons et al. 2013)
\begin{eqnarray}
ds^2 &=& \Theta\left(-\frac{\Delta}{r^2}
dt^2+\frac{r^2}{\Delta}dr^2+r^2d\vartheta^2\right) \nonumber \\
&& +\frac{r^2}{\Theta}(d\varphi-\Omega dt)^2\sin^2\vartheta,
\\
\Delta &=& r^2-2Mr+Q^2, \nonumber \\
\Theta &=& 1+\frac{1}{2}B^2(r^2\sin^2\vartheta+3Q^2\cos^2\vartheta) \nonumber \\
&& + \frac{1}{16}B^4(r^2\sin^2\vartheta+Q^2\cos^2\vartheta)^2,  \nonumber \\
\Omega &=&
-\frac{2}{r}QB+\frac{r}{2}QB^3\left(1+\frac{\Delta}{r^2}\cos^2\vartheta\right).
\nonumber
\end{eqnarray}
$M$ is the mass of the black hole, and $Q$ is the charge of the
black hole. $B$ stands for the strength of the magnetic field.
$\Omega$ is a dragging potential proportional to the coupling $QB$
because the interaction between the charge $Q$ and the magnetic
field $B$ serves as a rotating source for rotation; namely, it
directly arises from the charge. See also the paper of Santos $\&$
Herdeiro (2021) for more details on the metric.

This metric has two Killing vectors associated to stationarity and
axi-symmetry, which correspond to constant energy $E$ and angular
momentum $L$ of a test particle. The two constants satisfy the
relations
\begin{eqnarray}
\dot{t} &=& \frac{r^2}{\Theta\Delta}(E-\Omega L), \\
\dot{\varphi} &=&
\frac{r^2\Omega}{\Theta\Delta}E+\left(\frac{\Theta}{r^2\sin^2\vartheta}
-\frac{r^2\Omega^2}{\Theta\Delta}\right)L.
\end{eqnarray}
The spacetime determines the Hamiltonian
\begin{eqnarray}
H &=& H_1+\frac{\Delta p^2_r}{2r^2\Theta}+\frac{
p^2_\vartheta}{2r^2\Theta}, \\
H_1 &=& -\frac{r^2}{2\Theta\Delta}(E-\Omega L)^2+\frac{\Theta
L^2}{2r^2\sin^2\vartheta}.
\end{eqnarray}
If $Q=0$, the Hamiltonian (100) is nonintegrable. In this case,
chaos was shown by Li $\&$ Wu (2019). When $Q\neq 0$, the system
should also be nonintegrable.

Given the time transformation
\begin{eqnarray}
g(r,\vartheta)=\Theta,
\end{eqnarray}
the time-transformed Hamiltonian is
\begin{eqnarray}
\mathcal{H} &=& \mathcal{H}_1+\frac{\Delta p^2_r}{2r^2}+\frac{
p^2_\vartheta}{2r^2}, \\
\mathcal{H}_1 &=& p_0\Theta-\frac{r^2}{2\Delta}(E-\Omega
L)^2+\frac{\Theta^2 L^2}{2r^2\sin^2\vartheta}. \nonumber
\end{eqnarray}
The second term of Equation (103) contains three solvable parts,
as was shown by Wang et al. (2021b). Hence, the Hamiltonian (103)
has five solvable parts and the explicit symplectic schemes
(4)-(6) can work.

\subsubsection{Relativistic core-shell models}

Core-shell models describe black holes or neutron stars surrounded
by axially symmetric  shell of dipoles, quadrupoles, and
octopoles.  Vieira $\&$ Letelier (1999) gave these models in the
Schwarzschild coordinates $(t,r,\theta,\phi)$:
\begin{eqnarray}
ds^2 &=& -(1-\frac{2}{r})e^P dt^2
+e^{Q-P}[(1-\frac{2}{r})^{-1}dr^2 \nonumber \\
&& +r^2d\theta^2]+e^{-P}r^2\sin^2\theta d\phi^2,
\end{eqnarray}
where $Q$ and $P$  are two complicated functions of $r$ and
$\theta$ consisting of  multipoles.

We easily establish our explicit symplectic algorithms for the
obtained time transformation Hamiltonian by taking the time step
function
\begin{eqnarray}
g(r,\theta)=e^{Q-P}.
\end{eqnarray}

\subsubsection{Kerr-Newman solution with disformal parameter}

Let a disformal parameter $\beta$ describe the deviation of
modified vector tensor theory from the usual Einstein-Maxwell
gravity. The action of such a modified gravity can derive a
Kerr-Newman solution (Filippini $\&$ Tasinato 2018):
\begin{eqnarray}
ds^2 &=& -\frac{1}{\Sigma^2}(dt-a\sin^2\theta d\phi)^2
(\Delta\Sigma+\beta^2Q^2r^2) \nonumber \\
&& +\frac{\sin^2\theta}{\Sigma}[adt-(a^2+r^2)d\phi]^2
\nonumber \\
&& + \Sigma\left(\frac{\Sigma
dr^2}{\Delta\Sigma-\beta^2Q^2r^2}+d\theta^2\right)
\\
\Delta &=& a^2+r^2-2Mr+Q^2, \nonumber \\
\Sigma &=& r^2+a^2 \cos^2\theta. \nonumber
\end{eqnarray}
Note that $M$, $Q$ and $a$ are the black hole mass, charge and
spin, respectively. In addition, $t\in [0,\infty)$, $r\in
(0,\infty)$, $\theta\in (0,\pi)$, and $\phi\in [0,2\pi)$.

In the motion of a particle around the black hole, there are
constant energy $E$ and angular momentum $L$:
\begin{eqnarray}
\dot{t}&=&
\frac{Eg_{\phi\phi}+Lg_{t\phi}}{g^2_{t\phi}-g_{tt}g_{\phi\phi}},
\\
\dot{\phi}&=&
\frac{Eg_{t\phi}+Lg_{tt}}{g_{tt}g_{\phi\phi}-g^2_{t\phi}},
\end{eqnarray}
where $g_{tt}$, $g_{t\phi}$ and $g_{\phi\phi}$ are metric
components. The long expressions of $\dot{t}$ and $\dot{\phi}$ can
be found in the paper of Nazar et al. (2019). We obtain the
Hamiltonian
\begin{eqnarray}
H &=& H_1+\frac{(\Delta\Sigma-\beta^2Q^2r^2)p^2_r}{2\Sigma^2}
+\frac{p^2_\theta}{2\Sigma} \\
H_1 &=& -\frac{1}{2\Sigma^2}(\dot{t}-a\sin^2\theta \dot{\phi})^2
(\Delta\Sigma+\beta^2Q^2r^2) \nonumber \\
&& +\frac{\sin^2\theta}{2\Sigma}[a\dot{t}-(a^2+r^2)\dot{\phi}]^2.
\end{eqnarray}
Taking the time transformation function
\begin{eqnarray}
g(r,\theta)=\frac{\Sigma^2}{r^4},
\end{eqnarray}
we have the following time transformation Hamiltonian
\begin{eqnarray}
\mathcal{H} &=& \mathcal{H}_1+\frac{p^2_r}{2r^4}
(\Delta\Sigma-\beta^2Q^2r^2)
+\frac{\Sigma p^2_\theta}{2r^4}, \\
\mathcal{H}_1 &=& -\frac{1}{2r^4}(\dot{t}-a\sin^2\theta
\dot{\phi})^2
(\Delta\Sigma+\beta^2Q^2r^2) \nonumber \\
&&
+\frac{\Sigma\sin^2\theta}{2r^4}[a\dot{t}-(a^2+r^2)\dot{\phi}]^2
\nonumber \\
&& +p_0 \frac{\Sigma^2}{r^4}.
\end{eqnarray}
The second term of Equation (112) can be split into five
integrable parts:
\begin{eqnarray}
\mathcal{H}_2 &=& \frac{p^2_r}{2}, \\
\mathcal{H}_3 &=& -\frac{M}{r}p^2_r, \\
\mathcal{H}_4 &=& \frac{p^2_r}{2r^2}[a^2(1+\cos^2\theta) \nonumber \\
&& +Q^2(1-\beta^2)], \\
\mathcal{H}_5 &=& -\frac{Mp^2_r}{r^3}a^2\cos^2\theta, \\
\mathcal{H}_6 &=& \frac{p^2_r}{2r^4}a^2(a^2+Q^2)\cos^2\theta.
\end{eqnarray}

The third term of Equation (112) is $\Gamma=\Sigma
p^2_\theta/(2r^4)$. It seems to be simple, but is solved in
somewhat complicated way. The Hamiltonian $\Gamma$ is rewritten as
\begin{eqnarray}
\Gamma &=& \frac{p^2_\theta}{2r^4}(r^2+a^2\cos^2\theta) \nonumber \\
&=& \frac{p^2_\theta}{2r^4}[r^2+a^2(1-\sin^2\theta)]  \nonumber \\
&=& \frac{p^2_\theta}{2r^4}(r^2+a^2)+(-\frac{p^2_\theta}
{2r^4}a^2\sin^2\theta)  \nonumber \\
&=& \mathcal{H}_7+\mathcal{H}_8.
\end{eqnarray}
$\mathcal{H}_7$ is easily solved. Now, let us focus on solving
$\mathcal{H}_8$. This Hamiltonian has the canonical equations
\begin{eqnarray}
\frac{dr}{d\sigma} &=& 0, \\
\frac{d\theta}{d\sigma} &=& -\frac{p_\theta}
{r^4}a^2\sin^2\theta,  \\
\frac{dp_\theta}{d\sigma} &=& \frac{p^2_\theta} {r^4}a^2\sin\theta
\cos\theta, \\
\frac{dp_r}{d\sigma} &=& -\frac{2p^2_\theta} {r^5}a^2\sin^2\theta.
\end{eqnarray}
Their analytical solutions are explicit functions of the new time
$\sigma=\sigma_0+h$:
\begin{eqnarray}
c_1 &=& p_{\theta_0}\sin\theta_0, \\
c_2 &=& \tan \left(\frac{\theta_0}{2}\right), \\
r &=& r_0, \\
\theta &=& 2
\arctan\left(c_2e^{-\frac{c_1ha^2}{r^4_0}} \right),
\\
p_\theta &=& \frac{c_1}{\sin\theta}, \\
p_r &=& p_{r_0}-\frac{2h}{r^5_0}a^2c^2_1.
\end{eqnarray}
Here, $r_0$,  $\theta_0$, $p_{\theta_0}$ and $p_{r_0}$ are the
values of $r$, $\theta$, $p_{\theta}$ and $p_{r}$ at the new time
$\sigma_0$. Two problems are worth noting. Why is $\cos^2\theta$
replaced with its equivalent form $1-\sin^2\theta$ in Equation
(119)? If $\cos^2\theta$ is still used, $\sin\theta$ becomes
$\cos\theta$ in Equation (128). When $\theta=\pi/2$, the
computation of $c_1/\cos\theta$ does not continue. Why is
$\cos^2\theta$ not replaced with its another equivalent form
$[1+\cos(2\theta)]/2$ in Equation (119)? If it is, no explicitly
analytical solutions are given to the Hamiltonian $\Gamma$.

It is clear that the Hamiltonian (112) has 8 explicitly solvable
pieces. Thus, the Hamiltonian is typically suitable for the
application of the explicit symplectic methods (4)-(6). In such a
similar way, these constructions can be generalized to a rotating
non-Kerr black hole immersed in a uniform magnetic field
(Abdujabbarov et al. 2013). They are also applicable to a
non-axisymmetrical system of rotating black hole in external
magnetic field (Kop\'{a}\v{c}ek $\&$ Karas 2014), and a
modification to the Kerr-Newman black holes of general relativity
in Eddington-inspired Born-Infeld gravity (Guerrero et al. 2020).

Wu et al. (2021) confirmed that the fourth-order explicit
algorithm $S_4$ for the Kerr black hole is superior to the
fourth-order implicit symplectic method and the fourth-order
explicit and implicit mixed symplectic method  in computational
efficiency. The efficiency superiority of the application of the
explicit algorithms to the other black hole spacetimes should not
be altered.

\section{Summary}

Following the previous works (Wang et al. 2021a, 2021b, 2021c; Wu
et al. 2021; Sun et al. 2021a), we have developed explicit
sympelcetic algorithms for the long-term numerical integration of
orbits in general relativity and modified theories of gravity. We
mainly address one problem of which Hamiltonians of curved
spacetimes are directly split into multi explicitly integrable
terms. We also solve another problem of which Hamiltonians of
curved spacetimes are not but time transformation Hamiltonians of
curved spacetimes are. The key problem  how to split these
Hamiltonians or time transformation Hamiltonians is particularly
considered.

For the spacetimes given in Equation (8), their corresponding
Hamiltonians are directly split in the desirable forms and
natrurally allow for the application of explicit sympelcetic
integrators. Without loss of generality, these spacetimes include
the Schwarzschild black hole, Reissner-Nordstr\"{o}m anti de
Sitter black hole, Reissner-Nordstr\"{o}m-(de Sitter)-Anti-de
Sitter black hole surrounded by quintessence and a cloud of
strings and rotating black ring, etc. In particular, the
Hamiltonian of rotating black ring is shown to have 13 explicitly,
analytically solvable splitting parts.

However, the Hamiltonians of most metrics such as Equations (70)
and (89) are not directly separable into several explicitly
integrable pieces. Instead, appropriate time transformation
Hamiltonians to the Hamiltonians are. In this way, explicit
sympelcetic schemes are still available for these types of
spacetimes. The established symplectic algorithms use fixed time
steps in the new time, but might adopt adaptive time steps in the
original  proper time. Some of the spacetimes meeting this
requirement are the rotating black ring, regular black holes,
Gauss-Bonnet black hole, Kerr black hole, Majumdar-Papapetrou
dihole spacetime, Reissner-Nordstr\"{o}m-Melvin black holes,
core-shell models, and Kerr-Newman solution with disformal
parameter, etc. For example, an 8 part split  is given to the
time-transformed Hamiltonian of Kerr-Newman solution with
disformal parameter.

The splitting methods of the Hamiltonians or time-transformed
Hamiltonians associated to curved spacetimes  are not altered in
general when external magnetic fields surround the central bodies.
Although various splitting methods can be given to a Hamiltonian,
the number of splitting Hamiltonian pieces should be small as much
as possible so as to reduce roundoff errors. Many time
transformation functions can also be given to a Hamiltonian.

The multi part split  explicit symelectic integrators for
Equations (8), (70) and (89) are appropriate for most of the
spacetimes we have known. This brings a great extension to the
application of explicit symelectic methods for integrations of
orbits in curved spacetimes.  Such algorithms provide effective
means to numerically study various dynamical problems in general
relativity and modified theories of gravity. They are suited for
studying the transition from regular to chaotic dynamics of
charged test particles moving near black holes immersed in
external magnetic fields, such as a rotating black hole surrounded
by an external non-axisymmetrical magnetic field (Kop\'{a}\v{c}ek
$\&$ Karas 2014). Extreme-mass-ratio-inspirals are important
sources for the space-borne gravitational wave detectors. Their
orbits need to be integrated very accurately (Zhang $\&$ Han 2021;
Zhang et al. 2021), and these explicit symelectic methods may be
useful to this kind of dynamical systems. It should be good to use
the explicit symelectic integrators rather than non-symplectic
Runge-Kutta methods in ray-tracing codes on black hole shadows (Pu
et al. 2016).

\section*{Acknowledgments}

The authors are very grateful to a referee for valuable comments
and suggestions. This research has been supported by the National
Key R$\&$D Program of China (No. 2021YFC2203002), the National
Natural Science Foundation of China [Grant Nos. 11973020,
U2031145, 12173071], and the Natural Science Foundation of Guangxi
(Grant No. 2019JJD110006).

\end{document}